\documentclass[aps,prb,showpacs,superscriptaddress,amsmath,twocolumn]{revtex4-1}

\usepackage{amsmath}
\usepackage{amssymb}
\usepackage{graphicx}

\begin{document}

\title{Emergence of bound states in ballistic magnetotransport of graphene antidots}

\author{P. Rakyta}
\affiliation{Department of Theoretical Physics, Budapest University of Technology and Economics,
Condensed Matter Research Group of the Hungarian Academy of Sciences,
H-1111 Budafoki \'ut. 8, Hungary}

\author{E. T\'ov\'ari}
\author{M. Csontos}
\author{Sz. Csonka}
\affiliation{Department of Physics,
Budapest University of Technology and Economics,
Condensed Matter Research Group of the Hungarian Academy of Sciences,
H-1111 Budafoki \'ut. 8, Hungary}

\author{A. Csord\'as}
\affiliation{MTA-ELTE Theoretical Physics Research Group,
             P\'azm\'any P. S\'et\'any 1/A, H-1117 Budapest, Hungary}
\affiliation{Department of Physics of Complex Systems,
E{\"o}tv{\"o}s University,
H-1117 Budapest, P\'azm\'any P{\'e}ter s{\'e}t\'any 1/A, Hungary
}
\author{J. Cserti}
\affiliation{Department of Physics of Complex Systems,
E{\"o}tv{\"o}s University,
H-1117 Budapest, P\'azm\'any P{\'e}ter s{\'e}t\'any 1/A, Hungary
}


\begin{abstract}
An experimental method for detection of bound states around an antidot formed by a hole in a graphene sheet is proposed via measuring the ballistic two-terminal conductance. 
In particular, we consider the effect of bound states formed by a magnetic field on the two-terminal conductance and
 show that one can observe Breit-Wigner-like resonances in the conductance as a function of the Fermi level close to the energies of the bound states.
In addition, we develop a numerical method utilizing a reduced computational effort compared to the existing numerical recursive Green's function methods. 
\end{abstract}


\maketitle

\section{Introduction}

After the isolation of single-layer graphene flakes for the first time \cite{novoselov}, fabrication methods of graphene devices have undergone great development. It
turned out that the intrinsic charge traps in standard thermally grown SiO$_{2}$ substrates along with the residuals from the device preparation processes
place a major limitation on the mean free path. Novel fabrication procedures like suspending the graphene region\cite{stormer,largeyield} or using BN
substrates\cite{hBN} enabled the realization of high quality ballistic graphene devices above the micrometer scale.\cite{micrometer_ballistic,micrometer_ballistic2}
Using basic device geometries, several ballistic phenomena were recently demonstrated, where the
two-dimensionality and the Dirac spectrum of graphene play important roles: Due to Klein tunneling sharp Fabry-P\'erot interference was observed in $p-n$
junctions,\cite{pnrickhaus,pngrushina} conductance quantization was seen in narrow suspended graphene constrictions,\cite{quntumHall} and evidence of ballistic
trajectories was observed in BN-based\cite{focusing,cross1} and suspended devices.\cite{cross2} The high quality of these novel devices is well demonstrated by
observation of the quantum Hall effect (QHE) even at magnetic fields below 100 mT owing to the weak impurity potential.\cite{quntumHall,quntumHall2} These
fabrication techniques open new routes to experimental studies of various device geometries in the ballistic regime, thus making the theoretical
investigations of such systems also essential. 
Recent theoretical studies of periodic graphene antidot lattices (GALs) in the ballistic regime has predicted several possible experimental applications.\cite{transport_antidot,GAL,DOSinGAL,transport}
GALs are expected to be suitable for preparing resonant tunneling diodes\cite{transport_antidot} or graphene waveguides by a regular graphene strip surrounded by a GAL.\cite{GAL}

\begin{figure}[thb]
\includegraphics[scale=0.4]{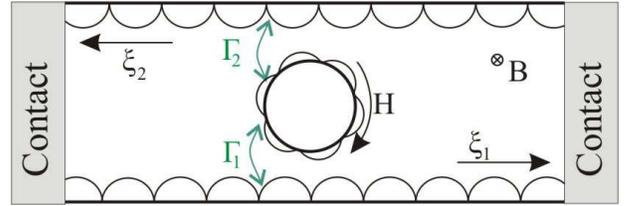}
\caption{(Color online) Graphene strip including a hole and contacted by metallic leads.
A magnetic field $B$ perpendicular to the sheet of graphene generates bound state $H$ localized to the hole and propagating edge states $\xi_1$ and $\xi_2$.
The edge states are coupled to the bound states due to scattering on impurities in the graphene lattice. The couplings are modeled by the hopping terms $\Gamma_1$ and $\Gamma_2$.
\label{fig:geometry}} 
\end{figure}

In this paper we argue that coherent ballistic transport can be used to study the properties of the bound states (BS's) in graphene nanostructures.
To this end, we studied the conductance of a graphene antidot attached to two metallic contacts as shown in Fig.~\ref{fig:geometry}.
Between the contacts propagating edge states $\xi_i$ ($i=1,2$) are formed by applying a homogeneous magnetic field perpendicular to the plane of the graphene sheet.
Moreover, due to the magnetic field the system also exhibits BSs (labeled by $H$ in Fig.~\ref{fig:geometry}) localized to the hole. 
We argue that resonant peaks can be observed in the conductance at energies close to the BS's even for small couplings between the edge states and the BSs. 
Moreover, such resonant peaks can be described by the well-known Breit-Wigner formula.\cite{BWresonance}
Similar anti resonance shapes were obtained in the conductance curves of graphene nanoribbons in the presence of vacancies that can be considered as a special case of antidots.\cite{minAntidot}
Later on we estimate the coupling terms $\Gamma_i$ ($i={1,2}$) which depend on the strength of the scattering impurities and on the decay length (DL) of the edge states. 
The edge states are localized to the edges and vanish exponentially with the distance measured from the edge on a length scale given by the DL.
In general the DL is governed by the magnetic length $l_B=\sqrt{\hbar/|eB|}$.
In the case of zigzag edges and an energy range close to the Dirac point, however, the edge states penetrate into the bulk on a length scale larger 
than $l_B$ due to the charge accumulation over the edge.\cite{iop}

The rest of the paper is organized as follows.
In Sec.~\ref{sec:BWresonance} we present an effective model predicting resonant peaks in the conductance as a function of the Fermi level.
In Sec.~\ref{sec:numres} we present our numerical results obtained by tight-binding (TB) calculations and compare them to the predictions of the effective model introduced 
in Sec.~\ref{sec:BWresonance}.
The details of our numerical approach are presented in Sec.~\ref{sec:num-calc}.
Finally we summarize our work in Sec.~\ref{sec:summary}.

\section{Breit-Wigner-like resonances in the conductance}  \label{sec:BWresonance}

In order to see the behavior of the conductance of a graphene nanostructure containing an antidot as shown in Fig.~\ref{fig:geometry} we develop an effective model and 
show that the conductance as a function of the Fermi level exhibits Breit-Wigner like resonances\cite{BWresonance} close to the BSs. 
The effective Hamiltonian of the antidot in a narrow energy window around a given BS of energy $E_{BS}$ in the basis $\left( \xi_1, BS, \xi_2 \right)^T$ can be written as
\begin{equation}
 H_{3}(E) = \left(\begin{matrix}
      E-g_1^{-1}     & \Gamma_1      & 0       \\
      \Gamma_1^*     & E_{BS}      & \Gamma_2^*    \\
      0          & \Gamma_2    & E-g_2^{-1}   
     \end{matrix}\right), \label{eq:H_WB}
\end{equation}
where $g_i$ is the Green's function of the edge state $\xi_i$.
For simplicity, we assume that the $g_i$'s are scalars depending weakly on the energy and are identical for the two edge states: $g_1=g_2=g$.
To calculate the transmission probability between the edge states $\xi_1$ and $\xi_2$ we first eliminate the degree of freedom related to the BS using the decimation method.\cite{decimation}
Then the reduced effective Hamiltonian becomes
\begin{subequations}
\begin{equation}
 H_{2}(E) = \left(\begin{matrix}
      E-g^{-1}+\gamma_{11}(E)     & \gamma_{12}(E)       \\
      \gamma_{21}(E)   & E-g^{-1} +\gamma_{22}(E) 
     \end{matrix}\right), \label{eq:H_WB2}
\end{equation}
where
\begin{equation}
 \gamma_{ij}(E) = \frac{\Gamma_i\Gamma_j^*}{E-E_{BS}}.
\end{equation}%
\label{effHam:eq}%
\end{subequations}%
The transmission probability $T_{2\rightarrow1}$ between the edge states is determined by the $G_{2}^{(12)}$ element of the Green's function $G_2 = \left[ E - H_2(E) \right]^{-1}$.
Thus, using Eq.~(\ref{effHam:eq}) the transmission probability can be computed analytically and we found
\begin{subequations}
\begin{equation}
 T_{2\rightarrow1}(E)\sim \left| G_{2}^{(12)}(E) \right|^2 = \frac{ \left|\Gamma_1\right|^2\left|\Gamma_2\right|^2}{ \left(E-E_{\rm res}\right)^2 + \Delta_0^2}, \label{eq:transmission}
\end{equation}
where
\begin{equation}
 E_{\rm res} = E_{BS}+\left(\left|\Gamma_1\right|^2+\left|\Gamma_2\right|^2\right){\rm Re}(g)\; \label{eq:resonantE}
\end{equation}
is the resonant energy and
\begin{equation}
 \Delta_0 = \left(\left|\Gamma_1\right|^2+\left|\Gamma_2\right|^2\right){\rm Im}(g) \label{eq:resonant_delta}
\end{equation}
\end{subequations}
stands for the width of the resonance.
As we can see from Eq.~(\ref{eq:transmission}), the transmission probability $T_{2\rightarrow1}(E)$ exhibits a Breit-Wigner resonance close to the $E_{\rm BS}$ energy of the BS.
Consequently, the conductance $C(E)\sim[1-T_{2\rightarrow1}(E)]$ between the metallic contacts also manifests resonances at the same energies as the transmission probability $T_{2\rightarrow1}(E)$.
The resonances, however, are shifted from $E_{\rm BS}$ by the self-energy given in Eq.~(\ref{eq:resonantE}).

In the low-density limit the $\Gamma_i$ coupling strengths can be approximated by first-order perturbation as $\Gamma_i\approx\langle\xi_i|V|\textrm{BS}\rangle$, 
where the scattering potential of the impurities is $V(\mathbf{r}) = \nu(\mathbf{r})\rho(\mathbf{r})$ [with the density $\rho(\mathbf{r})$ and 
the strength of the scattering impurities $\nu(\mathbf{r})$]. 
Edge states localized due to the magnetic field decay as $|\xi_i\rangle\sim e^{-x^2/(2l_B^2)}$, where $x$ is the distance measured from the edge.
Here the decay factor can be extracted from the asymptotic expansion of the wave function describing electronic states in a magnetic field as given in Ref.~\onlinecite{snake}.
In the case of a zigzag edge used in our numerical calculations, the charge accumulation over the edge\cite{iop} 
for energies close to the charge neutrality point increases the DL of the edge states.
If the distribution of the scattering impurities is homogeneous [$\rho(\mathbf{r})\equiv\rho$, $\nu(\mathbf{r})\equiv\nu$], the coupling strength becomes proportional to 
\begin{equation}
 \Gamma_i\sim\mathcal{O}(\nu\rho e^{-D^2/(4l_B^2)}), \label{eq:Gamma}
\end{equation}
where $D$ labels the hole-edge distance.
Note that the leading contribution to the $\Gamma_i$'s is related to impurities located at $x\sim D/2$.
As one can see, the coupling strengths $\Gamma_i$ are sensitive to the strength of the magnetic field and to the hole-edge distance.
At finite temperature the width of the resonance also depends on the temperature broadening $k_BT$.
Consequently, since $\Delta_0$ vanishes exponentially for $D/l_B\gtrsim\mathcal{O}(1)$ the width of the resonance will be governed by the temperature broadening rather 
than by $\Delta_0$ calculated at the zero-temperature limit.

\section{Numerical Results} \label{sec:numres}

To explore the behavior of the conductance of our graphene antidot nanostructure we calculated the conductance between two electrodes at zero temperature as a function of the Fermi level 
for the graphene antidot using the tight-binding method~\cite{wakabayashi} and interpreted the results using the effective model introduced in the previous section. 
The calculations were performed on graphene ribbons with zigzag edges and only the nearest-neighbor hopping $\gamma$ was taken into account.  
The details of our numerical approach including the improvement of the usual recursive Green's function method are discussed in Sec.~\ref{sec:num-calc}.

\begin{figure}[thb]
\includegraphics[scale=0.9]{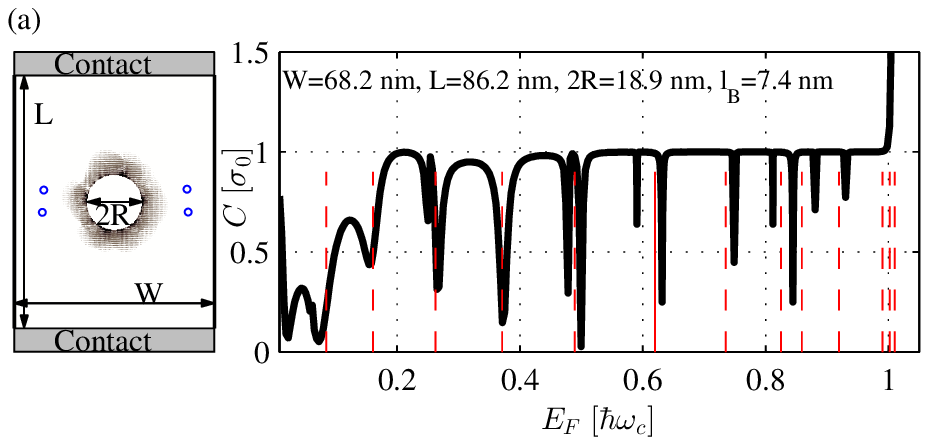}
\includegraphics[scale=0.9]{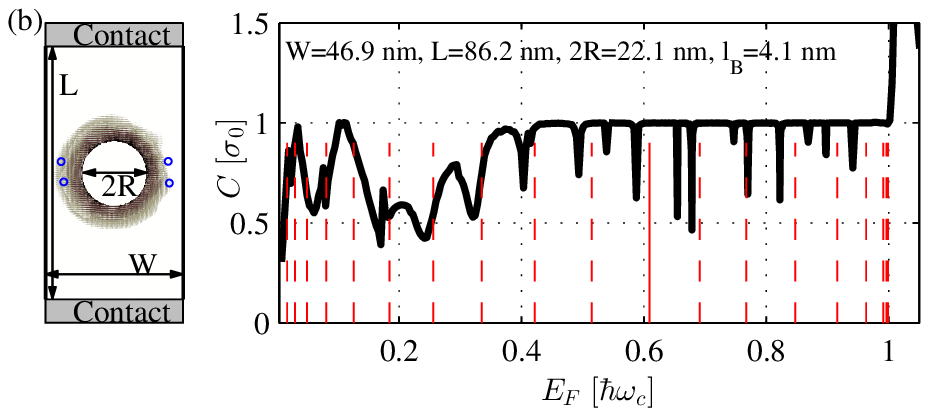}
\caption{(Color online) Left panels: Graphene strip of width $W$ and length $L$.
The hole in the strip has a diameter $2R$.
Blue circles correspond to atomic vacancies in the lattice.
The wave functions localized to the hole are also plotted in (a) and (b) at energies $E=0.619\;\hbar\omega_c$ and $E=0.606\;\hbar\omega_c$, respectively. 
\emph{Right panels}: The conductance (solid black line) in units of $\sigma_0 = 2e^2/h$ as a function of the Fermi energy. 
Vertical dashed and solid red lines correspond to BSs localized to the hole, where 
the solid line labels the energy of the BS plotted in the left panels.
Data in the figures were calculated at magnetic fields given by the magnetic length $l_B$ (see the numerical values in the subplots).
The magnetic flux in the hole is $\phi=\pi R^2B=0.817\phi_0$ ($\phi_0=h/e$ is the flux quantum) and $\phi=3.72\phi_0$ for (a) and (b), respectively.
The metallic contacts were modeled as heavily doped graphene leads shifted by a potential of $V = -0.27\gamma$ (where $\gamma$ is the coupling term between the nearest neighbor sites of the lattice).
\label{fig:cond-main}} 
\end{figure}

Our main results are summarized in Fig.~\ref{fig:cond-main}. 
The calculated conductance corresponding to a geometry where the radius of the hole is small compared to the width of the ribbon is shown in  Fig.~\ref{fig:cond-main}(a), 
while the opposite limit is shown in Fig.~\ref{fig:cond-main}(b).

According to Eq.~(\ref{eq:transmission}) the expected resonances are most pronounced if $\Gamma_1\approx\Gamma_2$.
To this end our calculations were performed on systems exhibiting an approximately symmetrical arrangement of the scattering centers around the hole that was located at the center of the strip.
The hole-edge distance and the magnetic field were chosen to meet the $D/l_B\gtrsim\mathcal{O}(1)$ condition in order to obtain well-separated resonances. 
As Fig.~\ref{fig:cond-main} shows one can indeed observe resonant peaks in the conductance at energies close to the BSs of the hole denoted by red vertical lines in the subplots of Fig.~\ref{fig:cond-main}.
The energy eigenvalues of the BSs are calculated within a TB framework discussed later on in this section.
Our numerical results indicate that the qualitative features of the conductance are not sensitive to the $R/W$ ratio (where $R$ is the radius of the hole and $W$ is the width of the strip).
In the calculations we consider only energies below the first Landau level $\hbar\omega_c=\sqrt{2|eB|\hbar v_F^2}$ (with the Fermi velocity $v_F$)  where there is only one propagating channel per edge.
The conductance calculated at higher energies (not shown in the figures) manifests a complex interference pattern, and hence we restrict our attention to the energy range $0<E<\hbar\omega_c$.
We study the energy dependence of the conductance considering two ranges.
First we examine the energy range where the DL of the edge states is smaller than the hole-edge distance $D$, while in the second energy range the DL becomes the larger length scale.

(i) For energies where the DL of the edge states is smaller than $D$, the coupling between the BSs and edge states can be increased by the presence of scattering impurities. 
According to our numerical results atomic vacancies are the most efficient scattering centers (see the blue circles in the left panels of Fig.~\ref{fig:cond-main}).
In the discussed energy range the conductance equals to one conductance unit $\sigma_0=2e^2/h$ (with the factor of $2$ standing for the spin degeneracy) corresponding to one open channel in the ribbon [see $E\gtrsim0.2\hbar\omega_c$ in Fig.~\ref{fig:cond-main}(a) and $E\gtrsim0.35\hbar\omega_c$ in Fig.~\ref{fig:cond-main}(b)].
The scattering centers affect the conductance only close to the energies corresponding to the BSs of the hole; otherwise the edge states do not backscatter to each other.
The model described in Sec.~\ref{sec:BWresonance} explains the shift between the resonances and the BSs.
However, at energies $\sim(0.5,0.6,0.85)\hbar\omega_c$ in Fig.~\ref{fig:cond-main}(a) and $\sim(0.5,0.7,0.75,0.85,0.9)\hbar\omega_c$ in Fig.~\ref{fig:cond-main}(b) one can see more than one resonance around each of the BSs which cannot be explained by this simple model.
In numerically exact calculations the couplings $\Gamma_i$ as well as the Green's functions $g_i$ are not scalars and might have stronger energy dependency hence the structure of the resonances becomes more complex as well.

(ii) At energies where the DL of the $\xi_i$ edge states is larger than $D$ (due to charge accumulation at the edge\cite{iop}) the backscattering of the edge states into each other is possible even without hitting a resonant energy state of the antidot.
Therefore the conductance over this energy range is smaller than $\sigma_0$ [see $E\lesssim0.2\hbar\omega_c$ in Fig.~\ref{fig:cond-main}(a) and $E\lesssim0.35\hbar\omega_c$ in Fig.~\ref{fig:cond-main}(b)].
\begin{figure}[thb]
\includegraphics[scale=0.7]{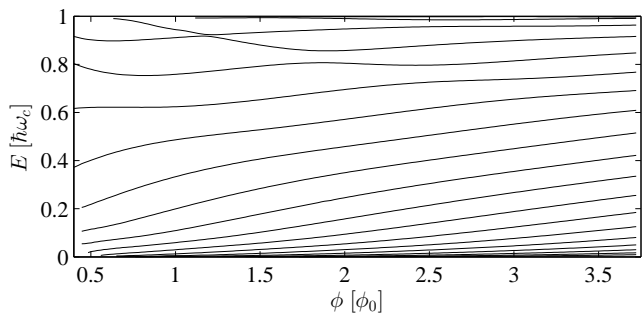}
\caption{Energy eigenvalues of BS's (in units of $\hbar\omega_c$) localized to the hole in the graphene strip as a function of the magnetic flux of the hole (in units of $\phi_0=h/e$), calculated 
for a hole with a diameter of \mbox{$2R=18.9$ nm}.
At low magnetic fluxes the lines tend to the closest Landau level, whereas at higher magnetic fields they become parallel to each other.
\label{fig:spectrum}} 
\end{figure}

Now it is clear that the observed resonances in the conductance are related to the BSs of the hole.
Therefore, we now describe the method to obtain the BSs localized to the hole and investigate the magnetic field dependence of these resonances.
In order to obtain the energy eigenvalues of the BSs we constructed the TB Hamiltonian of a strip including a hole of a given radius.
The Hamiltonian was constructed in the framework of the TB model described in Sec.~\ref{sec:num-calc} [see Eqs.~(\ref{eq:TB_ribbonHamiltonian})--(\ref{eq:vectorpot})], excluding sites located inside the hole.
Thus the hole inside the graphene strip was terminated by an atomically sharp edge where we did not assume any surface reconstruction of the lattice and the dangling bonds of the sites were not compensated either.
We numerically checked that the energy eigenvalues of the BSs are sensitive to the lattice termination over the perimeter of the hole, so one would find unique results for each antidot. 
The calculated wave functions of the BS's are also highly anisotropic due to the roughness of the lattice termination over the perimeter as shown in the left panels of Fig.~\ref{fig:cond-main}.
However, the qualitative properties of the calculated energy eigenvalues as a function of the magnetic field are robust. 
Since our goal here is to reveal physical properties of individual graphene antidots, we do not study the statistical properties of the BSs.
If the dimensions of the strip are large compared to the DL of the edge states $\xi_i$ the BSs become separated from the edge states since the BSs wave function exponentially vanishes close to the edges.
Thus, the energy eigenvalues of the BS's are robust to small variations in the dimensions of the strip which provides an efficient way to numerically separate BSs from the edge states.

Figure \ref{fig:spectrum} shows the energy eigenvalues of the BSs calculated for the hole corresponding to Fig.~\ref{fig:cond-main}(b) in the energy range of $0<E_n<\hbar\omega_c$ 
as a function of the $\phi$ magnetic flux inside the hole.
As one can see, the magnetic field dependence of the energy eigenvalues is nonmonotonic; however, at higher magnetic fields ($\phi\gtrsim3\phi_0$) the $E_n/\hbar\omega_c$ energy eigenvalues are expected to vary linearly with $\phi$.
In Fig.~\ref{fig:spectrum} one also can observe anticrossing points of the energy lines.
The anticrossings, even if they have no direct effect on the conductance, show interesting physical properties of the antidot.
Indeed, according to the calculations of Ref.~\onlinecite{antidot} based on the Dirac Hamiltonian, the valley degeneracy of the energy eigenvalues is lifted due to the boundary condition at the edge of the antidot.
Near atomically sharp scattering centers, such as edge terminations, the valley mixing in the electron states is enhanced due to scattering events with large momentum difference and results in the anticrossings of the energy lines.
For small magnetic fluxes the energy eigenvalues approach to the closest Landau level.
However, at low magnetic field it becomes difficult to calculate the energy eigenvalues of the BSs in the TB framework.
Since the DL of the wave functions becomes large for small magnetic fields, one needs to consider exceptionally large strips in order to separate the BSs from edge states.
Hence in Fig.~\ref{fig:spectrum} we calculated the energy eigenvalues of the BSs only for magnetic fluxes larger than $\sim0.5\phi_0$. 
Since the TB model discussed above manifests an electron-hole symmetry one can recover identical results in the energy range $-\hbar\omega_c<E_n<0$.  

Although in experiments the size of the samples is an order of magnitude larger than what we used in our numerical calculations, the obtained results are still relevant since  
we expect analogous physical properties for systems obeying the scaling law $(W,L,R)\rightarrow N\times(W,L,R)$ for the dimensions $B\rightarrow B/N^2$ ($l_B\rightarrow Nl_B$) 
for the magnetic field and $E_F\rightarrow E_F/N$ for the Fermi energy.
Thus, considering samples with a realistic size of $W\sim (0.6$--$1.0)\mu m$ and preserving the aspect ratio of Figs.~\ref{fig:cond-main}(a) and \ref{fig:cond-main}(b), we expect results similar to our calculations at magnetic fields \mbox{$B=(60$--$240)$ mT}.
According to the calculated energy eigenvalues of the BSs shown in Fig.~\ref{fig:spectrum} the average level spacing of the BSs at higher magnetic fields is about $\Delta E_{BS}\approx0.08\hbar\omega_c\propto\sqrt{B}$.
For a magnetic field in the range of $B=(60$--$240)$ mT the level spacing reads \mbox{$\Delta E_{BS}\approx(0.7$--$1.4)$ meV}.
In experiments one could reach the quantum Hall regime for suspended graphene samples at a temperature of $T\sim 4K$ and a magnetic field of $B\sim100$ mT.\cite{quntumHall,quntumHall2}
Since the temperature broadening \mbox{$k_BT\sim0.3$ meV} is smaller than $\Delta E_{BS}$ [and $\Delta_0$ in Eq.~(\ref{eq:Gamma}) vanishes exponentially with the magnetic field], it is feasible to observe the predicted resonances in the conductance in experiments, especially at higher magnetic fields or at lower temperatures than stated above.

\section{Details of the numerical calculations}
\label{sec:num-calc}

In this section we present the details of our numerical calculations for
the conductance shown in Fig.~\ref{fig:cond-main}. 
The transmission was calculated employing the Green's function technique of
Refs.~\onlinecite{sanvito} and \onlinecite{sanvito2} based on the nearest-neighbor TB model of graphene.\cite{wakabayashi}
However, in our approach we calculate the Green's function of the scattering region in a more efficient way than other Green's function techniques available in the literature.\cite{sanvito,sanvito2,recursiveGreen}
We construct the scattering region from a translationally invariant ribbon, and by using the Dyson equation\cite{Dyson} we can remove or add necessary
sites to the TB model in order to obtain the desired structure.
This procedure involves only sites that are directly related to the inhomogeneities of the scattering
region, and thus our approach is more efficient than the other Green's function methods, where the computations involve all sites of the scattering region. 

The electrodes in the calculations are assumed to be heavily doped, semi-infinite graphene nanoribbons. 
The nanoribbons and the scattering region including the hole are considered to be perfectly ballistic and the magnetic field is incorporated in the system by means of the Peierls substitution.\cite{peierls}

To understand the basic idea of our approach we first explain it on a scattering region including no hole or other scattering sites, and we set the value of the magnetic field to zero.

\begin{figure}[thb]
\includegraphics[scale=0.37]{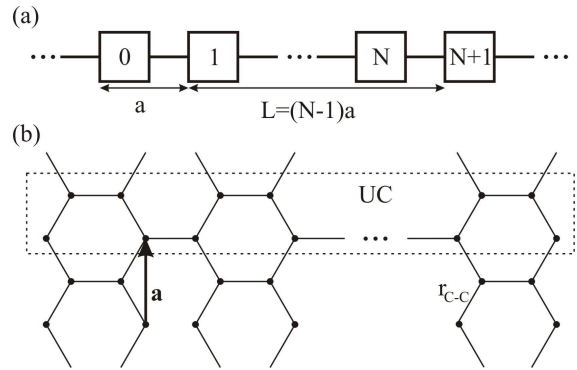}
\caption{(a) Schematic representation of an infinite ribbon made of unit cells coupled to each other by nearest-neighbor couplings.
The lattice constant is $a$ and the distance between the first and the $N$th unit cell is $L_N=(N-1)a$.
(b) The UC of a zigzag edged graphene ribbon is shown by a dashed rectangle. The translational vector is denoted by $\mathbf{a}$.
\label{fig:ribbon}} 
\end{figure}

\emph{(i) The minimal conductivity of graphene ($R=0$ and $B=0$)}.
\begin{figure}[thb]
\includegraphics[scale=0.7]{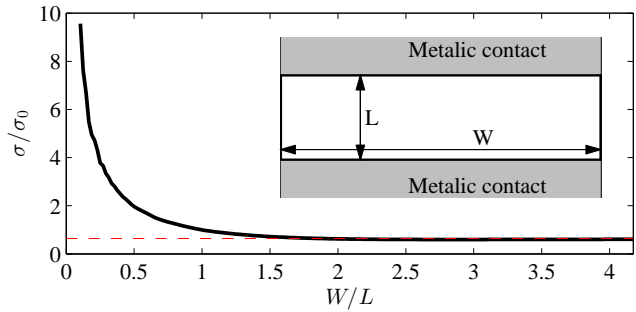}
\caption{(Color online) Solid black line: The calculated conductivity (in units of $\sigma_0 = 2e^2/h$) of a graphene strip with zigzag edges between contacts shown in the inset as a function of the aspect ratio $W/L$.
The length of the strip was \mbox{$L=20.4$ nm}, and the calculations were performed at the energy of $E=0.3\times10^{-4}\gamma$.
The metallic contacts were modeled as heavily doped graphene leads shifted by a potential of $V = -0.27\gamma$.
(Here $\gamma$ is the coupling term between the nearest-neighbor sites of the lattice.)
The theoretical value for the minimal conductivity of graphene is $\sigma_{\rm min}= 2\sigma_0/\pi$ (dashed red line).
\label{fig:mincond}} 
\end{figure}
To calculate the conductance of a graphene strip (see the inset of Fig.~\ref{fig:mincond}) one needs to calculate the Green's function of the strip. 
Then following the procedure described in Ref.~\onlinecite{sanvito} one can obtain the scattering matrix and the conductance of the strip attached to metallic contacts.
Note that the metallic contacts are coupled to the graphene strip via Dyson's equation, and hence the Green's function of the strip needs to be calculated only on its surface.
In order to calculate the surface Green's function of the strip, we start with the expression for the retarded Green's function $g_{zz'}$ of an infinite ribbon [see Eq.~(2.10) of Ref.~\onlinecite{sanvito}] made of identical unit cells (UCs) and coupled by nearest neighbor couplings (see Fig.~\ref{fig:ribbon}).
Indices $z$ and $z'$ label the UCs of the ribbon.
The Hamiltonian of the infinite ribbon is given by the block-diagonal matrix
\begin{equation}
 H_{TB} = \left(\begin{matrix}
	  \dots & \dots & \dots & \dots & \dots & \dots \\
	  \dots & H_0 & H_1 & 0 & 0 & \dots \\
	  \dots & H_1^{\dagger} & H_0 & H_1 & 0 & \dots\\
	  \dots & 0 & H_1^{\dagger} & H_0 & H_1 & \dots  \\
	  \dots & 0 & 0 & H_1^{\dagger} & H_0 & \dots		\\
	  \dots & \dots & \dots & \dots & \dots & \dots
	  \end{matrix}\right), \label{eq:TB_ribbonHamiltonian}
\end{equation}
where $H_0$ is the Hamiltonian of one UC:
\begin{equation}
 \left(H_0\right)_{ij} = \varepsilon_i\delta_{ij} - \gamma_{ij}(1-\delta_{ij}), \qquad \mathbf{R}_i,\mathbf{R}_j\in \textrm{UC}
\end{equation}
(with $\delta_{ij}$ the Kronecker delta), and the matrix $H_1$ contains the hopping amplitudes between nearest-neighbor UCs:
\begin{equation}
 \left(H_1\right)_{ij} = -\gamma_{ij}, \qquad \mathbf{R}_i,\mathbf{R}_j-\mathbf{a}\in \textrm{UC},
\end{equation}
where $\mathbf{a}$ is the translational vector of the ribbon (see Fig.~\ref{fig:ribbon}), $\mathbf{R}_i$ points to the $i$th site of the lattice, and $\varepsilon_i$ is the on-site energy.
In our calculations we take into account only nearest-neighbor hopping amplitudes $\gamma_{ij}\equiv\gamma$, where sites $i$ and $j$ denote nearest-neighbor carbon atoms in the honeycomb lattice for which $|\mathbf{R}_i-\mathbf{R}_j| = r_{\textrm{C-C}}$ [see Fig.~\ref{fig:ribbon}(b)].
In addition, the magnetic field can be incorporated by means of the Peierls substitution\cite{peierls}
\begin{equation}
 \gamma_{ij} = \gamma\;{\rm Exp}\left( \frac{2\pi{\rm i}}{\phi_0} \int\limits_{\mathbf{R}_j}^{\mathbf{R}_i} \mathbf{A(\mathbf{r})}{\rm d}\mathbf{r} \right) ,
\end{equation}
with the flux quantum $\phi_0 = h/e$ and the vector potential describing the magnetic field $\mathbf{B} = (0,0,B_z)^T$
\begin{equation}
 \mathbf{A}(\mathbf{r}) = \left[\left( \mathbf{r}\times\hat{\mathbf{a}}\right)_zB_z\right]\hat{\mathbf{a}}\;, \label{eq:vectorpot}
\end{equation}
where the $z$ direction is perpendicular to the plane of the ribbon and $\hat{\mathbf{a}}=\mathbf{a}/|\mathbf{a}|$.
Note that $\mathbf{A}(\mathbf{r})$ is translationally invariant along the ribbon, and hence the matrices $H_0$ and $H_1$ are identical for each UC.
Using Eq.~(2.10) of Ref.~\onlinecite{sanvito}, we can calculate the Green's function corresponding to UCs $z,z'\in\{0,1,N,N+1\}$ where the distance between UCs $1$ and $N$ is $L=(N-1)a$ (with $a$ the lattice constant), and UC $0$ ($N+1$) is the nearest-neighbor of UC $1$ ($N$) from the left (right).
The calculated Green's functions can be arranged into a matrix form as
\begin{equation}
 G = \left(\begin{matrix}
      g_{00}     & g_{01}      & g_{0N}     & g_{0(N+1)} \\
      g_{10}     & g_{11}      & g_{1N}     & g_{1(N+1)} \\
      g_{N0}     & g_{N1}      & g_{NN}     & g_{N(N+1)} \\
      g_{(N+1)0} & g_{(N+1)1} & g_{(N+1)N} & g_{(N+1)(N+1)}
     \end{matrix}\right). \label{eq:G}
\end{equation}
Since the structure of the ribbon contains only nearest neighbor hoppings between the UCs without long range interaction, the effective Hamiltonian defined as $H^{\rm eff} =EI - G^{-1}$ will have the following structure:
\begin{equation}
 H^{\rm eff} = \left(\begin{matrix}
      H_{00}     & H_{01}      & 0          & 0 \\
      H_{10}     & H_{11}      & H_{1N}     & 0 \\
      0          & H_{N1}      & H_{NN}     & H_{N(N+1)} \\
      0          & 0            & H_{(N+1)N} & H_{(N+1)(N+1)}
     \end{matrix}\right). \label{eq:Heff}
\end{equation}
Note that there is no effective coupling between UCs $0$ and $N$ since these UCs are coupled via UC $1$, and therefore the matrix element $H^{\rm eff}_{0N}$ vanishes.
For similar reasons the matrix elements $H^{\rm eff}_{N0}$, $H^{\rm eff}_{1(N+1)}$, $H^{\rm eff}_{N0}$, $H^{\rm eff}_{(N+1)0}$ and $H^{\rm eff}_{(N+1)1}$ also become zeros.
Using the Dyson's equation, let us apply a perturbation to the Hamiltonian $H^{\rm eff}$ given by
$V_1 = -H_{01}|0\rangle\langle 1| - H_{10}|1\rangle\langle 0|$ and $V_2 = -H_{N(N+1)}|N\rangle\langle N+1| - H_{(N+1)N}|N+1\rangle\langle N|$.
The potentials $V_1$ and $V_2$ uncouple a strip of length \mbox{$L_N$} from the rest of the ribbon. 
The perturbed Green's function of the ribbon defined by $\tilde{G}^{-1}=  EI-H^{\rm eff}-V_1-V_2$ will then fall apart into three separate subsystems:
\begin{equation}
 \tilde{G}= \left(\begin{matrix}
      g_L     & 0            & 0 \\
      0       & G_{\rm strip}          & 0 \\
      0       & 0         & g_R
     \end{matrix}\right), \label{eq:Gsubsystems}
\end{equation}
where $g_L$ ($g_R$) is the surface Green's function of the semi-infinite lead terminated by UC $0$ ($N+1$) from the right (left), and
\begin{equation}
 G_{\rm strip} = \left(EI - \left(\begin{matrix}
      H_{11}      & H_{1N} \\
      H_{N1}           & H_{NN}  
     \end{matrix}\right) \right)^{-1}
\end{equation}
is the surface Green's function of the strip terminated by UCs $1$ and $N$.
Note that the number of sites included in the calculations above is determined by the number of sites of one UC.
Therefore the computational cost of calculating the surface Green's function $G_{\rm strip}$ in the case of a graphene strip will be proportional to the width of the strip instead of its area.
We verified our numerical approach in two cases. 
First, the conductance calculations can be performed analytically on systems where the UC contains only one degree of freedom as in the case of one-dimensional chains.\cite{onedimchain}
Furthermore, we verify our numerical method by calculating the conductivity of a graphene strip 
(the conductivity is defined as $\sigma=C L/W$, where $L$ is the length, $W$ is the width, and $C$ is the conductance of the strip).

Figure \ref{fig:mincond} shows the calculated conductivity of a zigzag edged graphene strip as a function of the aspect ratio $W/L$.
As the aspect ratio becomes larger than $2$ the calculated conductivity tends to the value obtained for bulk graphene\cite{mincond} as demonstrated in Ref.~\onlinecite{CarloShotnoise}. 

\emph{(ii) Graphene antidot ($R>0$ and $B>0$)}. 
To calculate the surface Geen's function of a graphene antidot we follow the procedure described in the previous section.
First we calculate the Green's function $g_{zz'}$ at the sites of UCs $z,z'\in\{0,1,N,N+1,\{H\}\}$, where $\{H\}$ stands for UCs having finite overlap with the hole to be created.
The calculated Green's function then reads as
\begin{equation}
 G^{H} = \left(\begin{matrix}
      G     & g_{H,SP}      \\
      g_{SP,H}     & g_{H,H} 
     \end{matrix}\right),
\end{equation}
where $G$ is given by Eq.~(\ref{eq:G}), $g_{HH}$ is the Green's function calculated at sites $\{H\}$, and $g_{H,SP}$ and $g_{SP,H}$ are the propagators between sites $\{H\}$ and the surface points (SPs) of the strip.
Note that $g_{HH}$, $g_{H,SP}$, and $g_{SP,H}$ need to be calculated only at sites that are located inside the hole and have direct coupling to any sites outside the hole.
According to these conditions the calculations involve only sites that are close to the edge of the hole.
Hence the number of sites denoted by $\{H\}$ is proportional to the perimeter of the hole.
The second step of the approach described in the previous section is to calculate the effective Hamiltonian defined as
\begin{equation}
 H^{\rm eff,H} =EI - (G^H)^{-1} = \left(\begin{matrix}
      H_{SP,SP}     & H_{SP,H}      \\
      H_{H,SP}    & H_{H,H} 
     \end{matrix}\right),
\end{equation}
where the matrix $H_{SP,SP}$ has the same structure as $H^{\rm eff}$ given by Eq.~(\ref{eq:Heff}). 
Now we break all bonds between the hole and the rest of the ribbon by applying a perturbation to the Hamiltonian $H^{\rm eff,H}$ given by the potential 
$V_H = - H_{H,SP}|H\rangle\langle SP| - H_{SP,H}|SP\rangle\langle H|$.
We also apply the perturbations $V_1$ and $V_2$ in order to split the ribbon into three pieces as described in the previous subsection.
Finally one can extract the surface Green's function of the graphene antidot from the Green's function $\tilde{G}^H = \left( EI - H^{\rm eff,H} - V_1 - V_2 - V_H\right)^{-1}$ of the perturbed system as in Eq.~(\ref{eq:Gsubsystems}). 

Utilizing the current framework, one can also include scattering impurities in the calculations as atomic vacancies in the lattice which can be treated in the same way as the hole discussed above 
and/or scattering potentials that can be implemented via Dyson's equation. 
In the latter case the bonds coupling the perturbed sites to the rest of the lattice are not removed. 

The calculations described above involve a number of sites determined by the perimeter of the hole and by the width of the ribbon.
Therefore, the required computational effort of the approach scales linearly with the dimensions of the scattering region.

\section{Summary}
\label{sec:summary}

In this paper we showed that BSs generated around a graphene antidot by a perpendicular magnetic field can be detected by studying the two-terminal conductance in the ballistic regime. 
According to our calculations BSs localized to a hole in a graphene strip can be detected via the observation of Breit-Wigner-like resonances arising in the conductance as a function of the Fermi level.
The resonances can be observed close to the energies of the BSs.
The shift of the resonances measured from the energies of the BS's is determined by the coupling strength between the BSs and the edge states of the ribbon.
According to our estimates it is feasible to observe the appearance of the resonances in the conductance also in experiments.

In addition we also provided an efficient numerical method to calculate the surface Green's function of scattering regions containing several times $\sim 10^5$ sites.
The present numerical method, as an extension of the theoretical framework reported in Refs.~\onlinecite{sanvito,sanvito2}, involves only sites that are directly related to the inhomogeneities of the scattering region.
Thus our numerical approach is more efficient than other recursive Green's function techniques and can be utilized to study transport properties of other systems as well.

\section*{Acknowledgments}

We acknowledge support from EU ERC CooPairEnt 258789, FP7 SE2ND 271554, OTKA K105735.
P.~R., A.~Cs. and J.~Cs. would like to acknowledge the support of the Hungarian Scientific Research Fund No. K108676.
M.~C. and S.~C. are grantees of the Bolyai J\'anos Research Scholarship of the HAS.
M.~C. also acknowledges financial support from the European Union 7th Framework
Programme (Grant No.~293797).


\begin{thebibliography}{10}

\bibitem{novoselov} K.~S. Novoselov, 
A.~K. Geim, S.~V. Morozov, D.~Jiang, Y.~Zhang, S.~V.~Dubonos, I.~V.~Grigorieva, 
and A. A. Firsov, 
Science \textbf{306}, 666 (2004).

\bibitem{stormer} K.~I. Bolotin, K.~J. Sikes, Z. Jiang, M. Klima, G. Fudenberg, J. Hone, P. Kim, H. L. Stormer, Solid State Communications \textbf{146}, 351 (2008).

\bibitem{largeyield} N. Tombros, A. Veligura, J. Junesch, J. J. van den Berg, P. J. Zomer, M. Wojtaszek, I. J. V. Marun, H. T. Jonkman, B. J. van Wees, Journal of Applied Physics \textbf{109}, 093702 (2011).

\bibitem{hBN} C. R. Dean, A. F. Young, I. Meric, C. Lee, L. Wang, S. Sorgenfrei, K. Watanabe, T. Taniguchi, P. Kim, K. L. Shepard, J. Hone, Nature Nanotechnology \textbf{5}, 722 (2010)

\bibitem{micrometer_ballistic} V. E. Calado, S.-E. Zhu, S. Goswami, Q. Xu, K. Watanabe, T. Taniguchi, G. C. A. M. Janssen and L. M. K. Vandersypen, Appl. Phys. Lett. \textbf{104}, 023103 (2014).

\bibitem{micrometer_ballistic2} J. Baringhaus, M. Ruan, F. Edler, A. Tejeda, M. Sicot, A. Taleb-Ibrahimi, A.-P. Li, Z. Jiang, E. H. Conrad, C. Berger, C. Tegenkamp and W. A. de Heer, Nature \textbf{506}, 349-354 (2014).


\bibitem{pnrickhaus} P. Rickhaus, R. Maurand, M.-H. Liu, M. Weiss, K. Richter, C. Schoenenberger, Nature Comm. \textbf{4}, 2342 (2013).

\bibitem{pngrushina} A. L. Grushina, D. K. Ki, A. F. Morpurgo, Appl. Phys. Lett. \textbf{102}, 223102 (2013).

\bibitem{quntumHall}  N. Tombros, A. Veligura, J. Junesch, M. H. D. Guimaraes, I. J. Vera-Marun, H. T. Jonkman and B. J. van Wees, Nature Physics \textbf{7}, 697-700 (2011).

\bibitem{focusing} T. Taychatanapat, K. Watanabe, T. Taniguchi, P. Jarillo-Herrero, Nature Physics \textbf{9}, 225 (2013).

\bibitem{cross1} A. S. Mayorov, R. V. Gorbachev, S. V. Morozov, L. Britnell, R. Jalil, L. A. Ponomarenko, P. Blake, K. S. Novoselov, K. Watanabe, T. Taniguchi, A. K. Geim, Nano Letters \textbf{11}, 2396 (2011).

\bibitem{cross2} D.-K. Ki, A. F. Morpurgo, Nano Letters \textbf{13}, 5165 (2013).


\bibitem{quntumHall2}  D.~C. Elias, R.~V. Gorbachev, A. S. Mayorov, S. V. Morozov, A.~A. Zhukov, P. Blake, L. A. Ponomarenko, I. V. Grigorieva, K. S. Novoselov, F. Guinea and A. K. Geim,    Nature Physics \textbf{7}, 701-704  (2011).

\bibitem{transport_antidot}
T.~G. Pedersen and J. G. Pedersen, J. Appl. Phys. \textbf{112}, 113715 (2012).

\bibitem{GAL}
J. G. Pedersen, T. Gunst, T. Markussen, and T.~G. Pedersen, Phys. Rev. B. \textbf{86}, 245410 (2012).


\bibitem{DOSinGAL}
J.~G. Pedersen and T.~G. Pedersen, Phys. Rev. B. \textbf{87}, 235404 (2013).

\bibitem{transport}
M. Thomsen, S.~J. Brun, and T.~G. Pedersen, J. Phys.: Condens. Matter. \textbf{26}, 335301 (2014).

\bibitem{BWresonance} G. Breit and E. Wigner, Phys. Rev \textbf{49}, 519 (1936).

\bibitem{minAntidot}
D. A. Bahamon, A. L. C. Pereira, and P. A. Schulz, Phys. Rev. B \textbf{82}, 165438 (2010). 


\bibitem{iop} M. Arikawa, Y. Hatsugai and H. Aoki, J. Phys.: Conf. Ser. \textbf{150}, 022003 (2009).

\bibitem{decimation}
C.~J. Lambert, V.~C. Hui and S.~J. Robinson, J. Phys.: Condens. Matter \textbf{5}, 4187 (1993).

\bibitem{snake} 
L. Oroszl\'any, P. Rakyta, A. Korm\'anyos, C. J. Lambert, and J. Cserti, Phys. Rev. B \textbf{77}, 081403(R) (2008).


\bibitem{wakabayashi}
K.~Wakabayashi, M.~Fujita, H.~Ajiki, and M.~Sigrist,
Phys.~Rev.~B \textbf{59}, 8271 (1999).

\bibitem{antidot}
J.~G. Pedersen and T.~G. Pedersen, Phys. Rev. B \textbf{85}, 035413 (2012). 

\bibitem{sanvito}
S. Sanvito, C.~J. Lambert, J.~H. Jefferson and 
A. M. Bratkovsky, Phys. Rev. B \textbf{59}, 11936 (1999). 

\bibitem{sanvito2}
I. Rungger and S. Sanvito, Phys. Rev. B \textbf{78}, 035407 (2008). 

\bibitem{recursiveGreen}
C.~H. Lewenkopf and E.~R. Mucciolo , J. Comput. Electron. \textbf{12}, 203 (2013).

\bibitem{Dyson}
E.~N. Economou, \emph{Green's functions in Quantum Physics} p. 256, Springer-Verlag, Berlin (1983).

\bibitem{peierls} F. London, J. Phys. Radium \textbf{8}, 397 (1937).


\bibitem{onedimchain} 
 S. Datta, \emph{Quantum transport: atom to transistor} p. 213, Cambridge University Press, Cambridge (2005);
G. Cuniberti, G. Fagas, K. Richter, Chem. Phys. \textbf{281}, 465 (2002).


\bibitem{mincond} M.~I. Katsnelson, Eur. Phys. J. B \textbf{51}, 157-160 (2006); 
K. Ziegler, Phys. Rev. B \textbf{75}, 233407 (2007). 

\bibitem{CarloShotnoise}  J. Tworzydlo, B. Trauzettel, M. Titov, A. Rycerz, and C. W. J. Beenakker, Phys. Rev. Lett. \textbf{96}, 246802 (2006).





\end{thebibliography}
\end{document}